\documentclass[english,11pt]{revtex4-1}
\usepackage{graphicx}
\usepackage{hyperref}
\usepackage{cancel}
\usepackage{amssymb}
\usepackage{textcomp}
\usepackage{amsmath}
\usepackage{bm}
\usepackage{times}
\usepackage{epsfig}
\usepackage{color}
\usepackage{mathrsfs}
\usepackage{esint}
\usepackage{ulem}
\usepackage{subfigure}
\usepackage{makecell}
\usepackage{booktabs}

\makeatletter

\newcommand{\be}{\begin{equation}}
\newcommand{\ee}{\end{equation}}
\newcommand{\ba}{\begin{align}}
\newcommand{\ea}{\end{align}}
\def\bea{\begin{eqnarray}}
\def\eea{\end{eqnarray}}

\usepackage{slashed}

\makeatother

\usepackage{babel}

\newcommand{\hc}{{\rm h.c.}}

\newcommand{\eV}{{\rm eV}}

\newcommand{\GeV}{{\rm GeV}}
\newcommand{\TeV}{{\rm TeV}}

\begin{document}

\title{Radiative Seesaw Model and DAMPE Excess from Leptophilic Gauge Symmetry}

\author{Zhi-Long Han$^1$}
\email{sps\_hanzl@ujn.edu.cn}
\author{Weijian Wang$^2$}
\email{wjnwang96@aliyun.com}
\author{Ran Ding$^3$}
\email{dingran@mail.nankai.edu.cn}
\affiliation{
$^1$ School of Physics and Technology, University of Jinan, Jinan, Shandong 250022, China
\\
$^2$ Department of Physics, North China Electric Power University, Baoding 071003,China
\\
$^3$ Center for High Energy Physics, Peking University, Beijing 100871, China
}

\begin{abstract}
In the light of the $e^{+}+e^{-}$ excess observed by DAMPE experiment, we propose an anomaly-free radiative seesaw model with an alternative leptophilic $U(1)_X$ gauge symmetry. In the model, only right-handed leptons are charged under $U(1)_X$ symmetry. The tiny Dirac neutrino masses are generated at one-loop level and charged leptons acquire masses though the type-I seesaw-like mechanism with heavy intermediate fermions. In order to cancel the anomaly, irrational $U(1)_{X}$ charge numbers are assigned to some new particles.  After the spontaneous breaking of $U(1)_{X}$ symmetry, the dark $Z_{2}$ symmetry could appear as a residual symmetry such that the stability of inert particles with irrational charge numbers are guaranteed, naturally leading to stable DM candidates. We show that the Dirac fermion DM contained in the model can explain the DAMPE excess.  Meanwhile, experimental constraints from DM relic density, direct detection, LEP and anomalous magnetic moments are satisfied.
\end{abstract}

\maketitle
\section{Introduction}
It is well known that new physics beyond the Standard Model (SM) is needed to accommodate two open questions: the tiny neutrino masses and the cosmological dark matter (DM) candidates. The scotogenic model, proposed by Ma\cite{Ma:2006km,Ma:2007gq}, is one of the attractive candidate, which attributes the tiny neutrino masses to the radiative generation and the DM is naturally contained as intermediate messengers inside the loop. In the original models, an $\textsl{ad hoc}$ $Z_{2}$ or $Z_{3}$ symmetry serves to guarantee the stability of DM, whereas such discrete symmetry would be broken at high-scale \cite{Merle:2015gea}. Perhaps a more reasonable scenario is regarding the discrete symmetry as the residual symmetry originated from the  breaking of a continuous $U(1)$ symmetry at high scale. Along this line, several radiative neutrino mass models\cite{Adulpravitchai:2009re,Kanemura:2011vm,Chang:2011kv,
Kanemura:2011mw,Kanemura:2014rpa,Wang:2015saa,Ma:2016nnn,Ho:2016aye,Seto:2016pks,Wang:2017mcy, Nomura:2017emk,Nomura:2017vzp,Chao:2017rwv,Geng:2017foe,Nomura:2017kih,DeRomeri:2017oxa,Nanda:2017bmi,
Das:2017ski,Das:2017flq,Das:2017deo} were proposed based on gauged $U(1)_{B-L}$ theory, which is simplest and well-studied gauge extension of SM.

Very recently, a sharp excess in the $e^{+}+e^{-}$ flux is reported by Dark Matter Particle Explorer (DAMPE)~\cite{Ambrosi:2017wek}. If assuming a nearby subhalo locate at $0.1-0.3$ kpc away from the solar system, such excess can be interpreted by a 1.5 TeV DM particle with thermally averaged annihilation cross section $\sim10^{-26}~\text{cm}^3\text{s}^{-1}$ and predominately  annihilates into lepton final states~\cite{Yuan:2017ysv}. Inspired by this assumption, various models~\cite{Fan:2017sor,Fang:2017tvj,Duan:2017pkq,Gu:2017gle,
Athron:2017drj,Cao:2017ydw,Liu:2017rgs,Zu:2017dzm,Tang:2017lfb,Chao:2017yjg,Gu:2017bdw,Duan:2017qwj,
Zu:2017dzm,Tang:2017lfb,Huang:2017egk,Huang:2017egk,Jin:2017qcv,Gao:2017pym,Niu:2017hqe,Chao:2017emq,
Chen:2017tva, Li:2017tmd,Zhu:2017tvk,Gu:2017lir,Nomura:2017ohi,Ghorbani:2017cey,Cao:2017sju,Yang:2017cjm,
Ding:2017jdr,Liu:2017obm,Ge:2017tkd,Zhao:2017nrt,Sui:2017qra,Okada:2017pgr,Fowlie:2017abc,Cao:2017gsw} have been proposed. On the other hand, the model with $U(1)_{B-L}$ gauge extension is disfavored since it also predicts accompanying antiproton excess which is absent.

In this work,  we present a radiative neutrino mass model based on an alternative leptophilic $U(1)_{X}$ gauge symmetry. In the model only the right-handed SM leptons are charged under the $U(1)_X$ symmetry, resulting in the direct Yukawa couplings forbidden in the lepton sector.  We will show that the Dirac neutrino masses are generated radiatively and the charged leptons,  acquire masses via seesaw-like mechanism.  The heavy fermions we added for anomaly-free cancellation play as the intermediate fermions in lepton mass generation.  After the spontaneous symmetry breaking(SSB) of $U(1)_{X}$, the dark $Z_{2}$ symmetry could appear as a residual symmetry \cite{Krauss:1988zc} such that the stability of a classes of inert particles are protected by the irrational $U(1)_{X}$ charge assignments from decaying into SM particles, naturally leading to stable DM candidates.

The rest of this paper is organised as follows. In Sec.~\ref{Sec:MD},
the model is set up. In Sec.~\ref{Sec:DP}, we focus on DM phenomenon and its implication on DAMPE results. Conclusions are given in Sec.~\ref{Sec:CL}.

\section{Model Setup}\label{Sec:MD}
\subsection{Particle content and anomaly cancellation}
All the field contents and their charge assignments under $SU(2)_{L}\times U(1)_{Y}\times U(1)_{X}$ gauge symmetry are summarized in Table~\ref{TB:Charge}.  First of all, the $U(1)_{X}$ symmetry proposed here is a right-handed leptophilic gauge symmetry since, in SM sector, only right-handed leptons carry $U(1)_{X}$ charges.  As a result, the SM Yukawa coupling $\bar {L} \Phi E_{R}$ for charged lepton mass generation is strictly forbidden. In order to generate the minimal Dirac neutrino mass, two right-handed neutrino field $\nu_{Ri}(i=1,2)$ are added being coupled with $U(1)_X$, leading to the zero mass for the lightest neutrino. The direct Yukawa coupling $\bar{L}\widetilde{\Phi}\nu_{R}$ is also forbidden. Instead we have introduce several Dirac fermions as the intermediated fields for lepton mass generation with their corresponding chiral components $\Psi_{Ri/Li} (i = 1-9)$ and $F_{Ri/Li} (i = 1-4)$ respectively. In the scalar sector, we further add inert doublet scalars $\eta_1$, $\eta_2$ and one inert
singlet scalar $\chi$. An SM singlet scalar $\sigma$ is added being responsible for $U(1)_{X}$ breaking.

First and foremost, we check the anomaly cancellations for the new gauge symmetry in
the model. The $[SU(3)_C]^2U(1)_X$ and $[SU(2)_L]^2U(1)_X$ anomalies are zero because quark and left-handed leptons are not assumed coupled to $U(1)_{X}$. We then find all other anomalies are also zero because
\begin{eqnarray}
[U(1)_{Y}]^2U(1)_{X} &:& -3\times(-1)^2\times 3n+9\times(-1)^2\times 2n-9\times (-1)^2\times n=0
\\ \nonumber
[U(1)_{X}]^2U(1)_{Y} &:&  -3\times (3n)^2\times(-1)+9\times (2n)^2\times (-1)-9\times n^{2}\times (-1)=0
\\ \nonumber
[\text{Gravity}]^2 U(1)_{Y} &:&     Str[Q_{Y}]_{SM} +(-1)-(-1)=0
\\ \nonumber
U(1)_{Y}^{3} &:&  Str[Q_{Y}^3]_{SM}+(-1)^3-(-1)^3=0
\\ \nonumber
[\text{Gravity}]^2 U(1)_{X} &:&     -3\times 3n-2\times 2n +9\times 2n-9\times n
+ 4 \times Q_{F_L} - 4 \times Q_{F_R}=0
\\ \nonumber
[U(1)_{X}]^3 &:&  -3\times (3n)^3\!-2\times (2n)^3 \!+9\times (2n)^3\!-9\times n^3
\!+ 4 \times Q^3_{F_L} \! - 4 \times Q^3_{F_R} =0
\end{eqnarray}
Here, $Q_{F_L}=(\sqrt{11}+1)n/2$ and $Q_{F_R}=(\sqrt{11}-1)n/2$ are the $U(1)_X$ charge of $F_L$ and $F_R$ as shown in Table \ref{TB:Charge} respectively.
In order to cancel anomaly,  the $F_{Li/Ri}$ fermions acquire irrational $U(1)_X$ charge numbers. Similar scenarios also appeared in radiative inverse or linear seesaw models\cite{Kanemura:2014rpa,Wang:2015saa} where other solutions to anomaly free conditions with irrational $B-L$ charges of mirror fermions were found.

\begin{widetext}
\begin{center}
\begin{table}
\begin{tabular}{|c||c|c|c|c|c|c|c||c|c|c|c|c|c|}\hline\hline
&\multicolumn{7}{c||}{Lepton Fields} & \multicolumn{5}{c|}{Scalar Fields} \\\hline
& ~$L_L$~ & ~$E_R^{}$~& ~$\nu_{Ri}$ ~ &~$\Psi_{Lm}$ ~ &~$\Psi_{Rm}$ ~&~$F_{L\alpha}$ ~&~$F_{R\alpha}$ ~& ~$\Phi$ ~ & ~$\eta_{1}$~  & ~$\eta_{2}$ ~& ~$\chi$~  & ~$\sigma$ \\\hline
$SU(2)_L$& ~$2$~ & ~$1$~& ~$1$ ~ &~$1$ ~ &~$1$ ~&~$1$ ~&~$1$ ~& ~$2$ ~ & ~$2$~  & ~$2$ ~& ~$1$~  & ~$1$ \\\hline
$U(1)_Y$& ~$-\frac{1}{2}$~ & ~$-1$~& ~$0$ ~ &~$-1$ ~ &~$-1$ ~&~$0$ ~&~$0$ ~& ~$\frac{1}{2}$ ~ & ~$\frac{1}{2}$ ~& ~$\frac{1}{2}$~  & ~$0$~  & ~$0$ \\\hline
$U(1)_X$& ~$0$~ & ~$3n$~& ~$2n$ ~ &~$2n$ ~ &~$n$ ~&~$\frac{\sqrt{11}+1}{2}n$ ~&~$\frac{\sqrt{11}-1}{2}n$ ~& ~$0$ ~ & ~$n$~ & ~$\frac{\sqrt{11}-1}{2}n$ ~ & ~$\frac{\sqrt{11}-3}{2}n$~  & ~$n$ \\\hline
\end{tabular}
\caption{Contents of relevant particle fields. We have set $i=1,2$; $m=1-9$ and $\alpha=1- 4$  to satisfy the anomaly free condition. }
\label{TB:Charge}
\end{table}
\end{center}
\end{widetext}

\subsection{Scalar Sector}
The scalar potential in our model is given by
\begin{equation}\begin{split}
V&=\mu_{\Phi}^{2}\Phi^{\dag}\Phi +\mu_{\eta_{1}}^{2}\eta_{1}^{\dag}\eta_{1} +\mu_{\eta_{2}}^{2}\eta_{2}^{\dag}\eta_{2}+\mu_{\chi}^{2}|\chi|^{2}+\mu_{\sigma}^{2}|\sigma|^{2}
+\lambda_{\Phi}(\Phi^{\dag}\Phi)^{2}+\lambda_{\eta_{1}}(\eta_{1}^{\dag}\eta_{1})^{2}\\
&+\lambda_{\eta_{2}}(\eta_{2}^{\dag}\eta_{2})^{2}+\lambda_{\chi}|\chi|^{4}+\lambda_{\sigma}|\sigma|^{4}
+\lambda_{\eta_{1}\Phi}(\Phi^{\dag}\Phi)(\eta_{1}^{\dag}\eta_{1})
+\lambda^{\prime}_{\eta_{1}\Phi}(\eta_{1}^{\dag}\Phi)(\Phi^{\dag}\eta_{1})\\
& +\lambda_{\eta_{2}\Phi}(\Phi^{\dag}\Phi)(\eta_{2}^{\dag}\eta_{2})
+\lambda^{\prime}_{\eta_{2}\Phi}(\eta_{2}^{\dag}\Phi)(\Phi^{\dag}\eta_{2})
+\lambda_{\chi\Phi}|\chi|^{2}(\Phi^{\dag}\Phi)+\lambda_{\sigma\Phi}|\sigma|^{2}(\Phi^{\dag}\Phi) \\
&+\lambda_{\eta_{1}\eta_2}(\eta_1^{\dag}\eta_1)(\eta_{2}^{\dag}\eta_{2})
+\lambda^{\prime}_{\eta_{1}\eta_2}(\eta_{1}^{\dag}\eta_2)(\eta_2^{\dag}\eta_{1})
+\lambda_{\chi\eta_{1}}|\chi|^{2}(\eta_{1}^{\dag}\eta_{1})
+ \lambda_{\sigma\eta_{1}}|\sigma|^{2}(\eta_{1}^{\dag}\eta_{1})
\\
& +\lambda_{\chi\eta_{2}}|\chi|^{2}(\eta_{2}^{\dag}\eta_{2})\!+\! \lambda_{\sigma\eta_{2}}|\sigma|^{2}(\eta_{2}^{\dag}\eta_{2})
\!+\!\lambda_{\chi\sigma}|\chi|^{2}|\sigma|^{2}\!+\![\mu(\eta_{1}^{\dagger}\Phi)\sigma
\!+\!\lambda(\eta_{2}^{\dagger}\Phi)\chi\sigma+\hc]
\label{sc}\end{split}\end{equation}
The scalars $\Phi$ and $\sigma$ and $\eta_{1}$ with their vevs after SSB of $U(1)_{X}$ can be parameterized as
\begin{equation}
\Phi=\left(\begin{array}{c}
  G^{+}_{\phi}\\
  \frac{v_{\phi}+\phi^{0}+iG_{\phi}}{\sqrt{2}}
  \end{array}\right),\quad\quad
  \eta_{1}=\left(\begin{array}{c}
  \eta_{1}^{+}\\
  \frac{u}{\sqrt{2}}+\eta_{1}^0
  \end{array}\right),\quad\quad
  \sigma=\frac{v_{\sigma}+\sigma_{0}+iG_{\sigma}}{\sqrt{2}}.
 \end{equation}
Then the minimum of $V$ is determined by
\begin{equation}\begin{split}
v_{\phi}(\mu_{\Phi}^2+\lambda_{\Phi}v_{\phi}^2+\frac{1}{2}\lambda_{\sigma\Phi}v_{\sigma}^2
+\frac{1}{2}(\lambda_{\eta_{1}\Phi}+\lambda_{\eta_{1}\Phi}^{\prime})u^2)+\frac{\mu v_{\sigma}u}{\sqrt{2}}=&0,\\
v_{\sigma}(\mu_{\sigma}^2+\lambda_{\sigma}v_{\sigma}^2+\frac{1}{2}\lambda_{\sigma\Phi}v_{\phi}^2
+\frac{1}{2}\lambda_{\sigma\eta_{1}}u^2)+\frac{\mu v_{\phi}u}{\sqrt{2}}=&0,\\
u(\mu_{\eta_{1}}^2+\lambda_{\eta_1}u^2+\frac{1}{2}\lambda_{\sigma\eta_{1}}v_{\sigma}^2
+\frac{1}{2}(\lambda_{\eta_{1}\Phi}+\lambda_{\eta_{1}\Phi}^\prime)v_{\phi}^2)+\frac{\mu v_{\phi}v_{\sigma}}{\sqrt{2}}=&0.
\end{split}\end{equation}
For a large and negative $\mu_{\sigma}^2$, there exists a solution with $u^2\ll v_{\phi}^2\ll v_{\sigma}^2$ as
\begin{equation}
v_{\sigma}^2 \backsimeq\frac{-\mu_{\sigma}^2}{\lambda_\sigma},\quad v_{\phi}^2\backsimeq \frac{-2\mu_{\Phi}^2-\lambda_{\sigma\Phi}v_{\sigma}^2}{2\lambda_{\Phi}},\quad u^2 \backsimeq \frac{-\sqrt{2}\mu v_{\phi}v_{\sigma}}{2\mu_{\eta_1}^2+\lambda_{\sigma\eta_{1}}v_{\sigma}^2},
\end{equation}
where $v_{\phi}\simeq246$GeV is the vev of the SM higgs doublet
scalar and $v_{\sigma}$ is responsible for the SSB of $U(1)_{X}$ symmetry.  We have set a positive $\mu_{\eta_{1}}^2$, hence the vev of $\eta_{1}$ scalar is not directly acquired as that of $\Phi$ and $\sigma$ but induced from $\mu(\eta_{1}^{\dagger}\Phi)\sigma$ term. Note that $\eta_{2}$ and $\chi$ do not acquire vevs because of positive $\mu_{\eta_2}^2$, $\mu_{\chi}^2$ and the absence of linear terms, like $\chi\sigma^{k}$.

The mass spectrum of scalar $\sigma, \Phi$ and $\eta_{1}$ can be obtained with the  their vevs and the cross terms in Eq.\eqref{sc}. In the condition of $u^2\ll v_{\phi}^2\ll v_{\sigma}^2$, the contributions from $v_{\Phi}$ and $v_{\sigma}$ to scalar masses are dominant, and mixings between $\eta_1$ and other CP-even scalars are negligible small.
Then the two CP-even scalars $h$ and $H$ with mass eigenvalues are given by
\begin{eqnarray}
m_{h,H}^{2}\backsimeq\lambda_{\Phi}v_{\phi}^{2}+\lambda_{\sigma}v_{\sigma}^{2}\mp\sqrt{(\lambda_\Phi
v_{\phi}^{2}-\lambda_{\sigma}v_{\sigma}^{2})^{2}+\lambda_{\sigma\Phi}^{2}v_{\phi}^{2}v_{\sigma}^{2}},
\end{eqnarray}
with the mixing angle
\begin{equation}
\tan 2 \alpha = \frac{ \lambda_{\sigma\Phi} v_\Phi v_\sigma}{\lambda_{\sigma}v_{\sigma}^{2}-\lambda_\Phi
v_{\phi}^{2}},
\end{equation}
where we take scalar $h$ as the SM-like Higgs boson and $H$ the heavy Higgs boson.
A small mixing angle $\sin\alpha\sim0.1$ is assumed to satisfy Higgs measurement \cite{Khachatryan:2016vau}.
Note that due to the lack of $(\Phi^\dag\eta_{1,2})^2$ term, we actually have nearly degenerate masses for the real and imaginary part of $\eta_{1,2}^0$ \cite{Wang:2016vfj}, and they are assumed to be degenerate for simplicity in the following discussion.
The masses of scalar doublet $\eta_1$ are
\begin{eqnarray}
m_{\eta_{1}^0}^2&\backsimeq&\mu_{\eta_1}^2+\frac{1}{2}(\lambda_{\phi\eta_1}
+\lambda_{\phi\eta_{1}}^{\prime})v_{\phi}^{2}+\frac{1}{2}\lambda_{\sigma\eta_1}v_{\sigma}^{2},\\
m_{\eta_1^\pm}^2&\backsimeq&\mu_{\eta_1}^2+\frac{1}{2}\lambda_{\phi\eta_1}v_{\phi}^{2}
+\frac{1}{2}\lambda_{\sigma\eta_1}v_{\sigma}^{2}.
\end{eqnarray}
Hereafter, we take degenerate $\eta_1$ scalars and $m_{H,\eta_1}\sim500~\GeV$ for illustration.
For a complete detail of mass spectrum of $\Phi, \sigma, \eta_{1}$ scalars, one can refer some models which shares part of the scalar potential, e.g. Ref.\cite{Wang:2016vfj}.
On the other hand, we pay more attention to inert scalars $\eta_{2}$ and $\chi$ which are closely related to neutrino mass generation and DM.
Note that scalars $\eta_{2},\chi$ do not mix with $\Phi$, $ \sigma$ and $\eta_{1}$.
The two mass eigenstate of neutral complex scalars $\eta_{2}^0$ and $\chi$ are obtained by
\begin{equation}
 \left(\begin{array}{c}
  S_{1}\\
  S_{2}
  \end{array}\right)=\left(\begin{array}{cc}
  \cos\theta&-\sin\theta\\
  \sin\theta&\cos\theta
  \end{array}\right)\left(\begin{array}{c}
  \eta_2^{0}\\
  \chi
  \end{array}\right),\quad \sin 2\theta=\frac{\lambda v_{\phi}v_{\sigma}}{m_{S_{1}}^{2}-m_{S_{2}}^2},
\end{equation}
with mass eigenvalues
\begin{equation}
m_{S_{1,2}}^{2}=\frac{1}{2}\Big(M_{\eta_2}^2+M_{\chi}^2\pm \sqrt{(M_{\eta_2}^2-M_{\chi})^2+\lambda^{2}v_{\phi}^2 v_{\sigma}^2}
\Big),
\end{equation}
where
\begin{equation}\begin{split}
&M_{\eta_2}^2\simeq\mu_{\eta_2}^{2}+\frac{1}{2}(\lambda_{\phi\eta_2}
+\lambda_{\phi\eta_{2}}^{\prime})v_{\phi}^{2}
+\frac{1}{2}\lambda_{\sigma\eta_2}v_{\sigma}^{2},\\
&M_{\chi}^2\simeq\mu_{\chi}^{2}+\frac{1}{2}\lambda_{\chi\phi}v_{\phi}^{2}
+\frac{1}{2}\lambda_{\chi\sigma}v_{\sigma}^{2}.\\
\end{split}\end{equation}
Meanwhile, the mass of inert charged scalar $\eta_2^\pm$ is
\begin{equation}
M_{\eta_2^\pm}^2\simeq\mu_{\eta_2}^{2}+\frac{1}{2}\lambda_{\phi\eta_2}v_{\phi}^{2}
+\frac{1}{2}\lambda_{\sigma\eta_2}v_{\sigma}^{2}
\end{equation}
As will shown in Sec.~\ref{Sec:DP}, the DAMPE excess favors fermion DM $m_{F_1}\sim1.5~\TeV$. Therefore, heavier inert scalars, e.g., $m_{S_1,S_2,\eta_2^\pm}\sim10~\TeV$, are assumed.

\subsection{Lepton Masses}

\begin{figure}[!htbp]
\begin{center}
\includegraphics[width=0.5\linewidth]{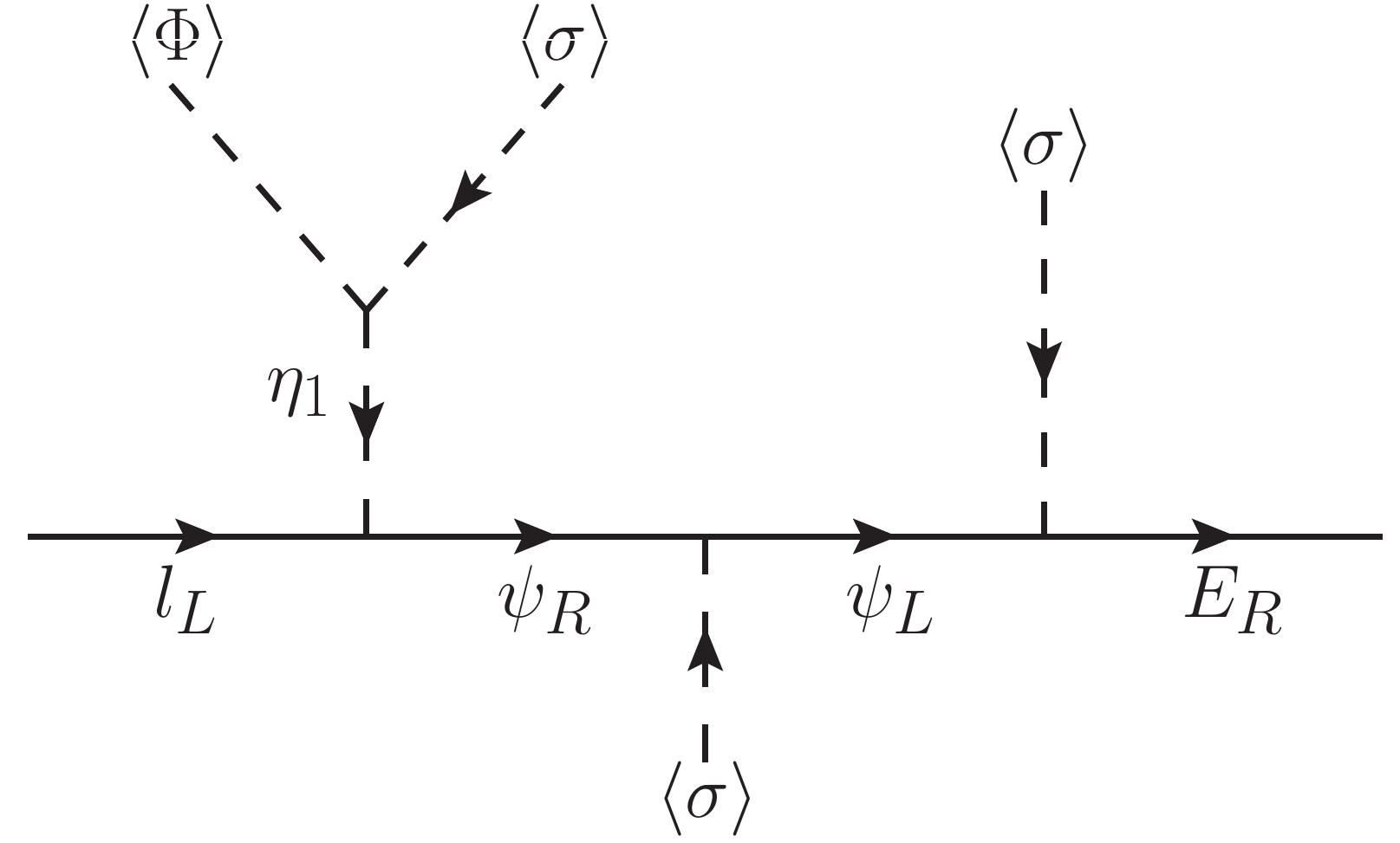}
\end{center}
\caption{ Charged lepton mass generation \label{lep}}
\end{figure}

The Yukawa interactions related to charged lepton mass generation is given by
\begin{equation}
\mathcal{L}_{1} ~\supset~ y_{1}\bar{L}\eta_{1}\Psi_{R}+y_{2}\bar{E}_{R}\Psi_{L}\sigma+y\bar{\Psi}_{L}\Psi_{R}\sigma+\hc
\label{y1}\end{equation}
the charged lepton masses are generated though the diagram in Fig. \ref{lep}. In the basis of $(\bar {l}_{L}, \bar{\Psi}_{L})$ and $(E_{R}, \Psi_{R})$, we obtain the $12\times 12$ effective mass matrix
\begin{equation}
\left(\begin{array}{cc}
  \bar {l}_{L}&\bar{\Psi}_{L}
  \end{array}\right)
\left(\begin{array}{cc}
  0&\frac{y_1 u}{\sqrt{2}}\\
  \frac{y_2 v_\sigma}{\sqrt{2}}&\frac{y v_{\sigma}}{\sqrt{2}}
  \end{array}\right)
  \left(\begin{array}{c}
  E_{R}\\
  \Psi_{R}
  \end{array}\right)+\hc
\end{equation}
Then the charged lepton mass is obtained as $M_{l}\simeq y_1 y_2  u/(\sqrt{2}y)$.
Correct charged lepton mass can be acquired with $y_{1,2}^e=8.5\times10^{-4}$, $y_{1,2}^\mu=1.2\times10^{-2}$ and $y_{1,2}^\tau=5.0\times10^{-2}$ for $u=10~\GeV$ and $y=0.01$.

\begin{figure}[!htbp]
\begin{center}
\includegraphics[width=0.4\linewidth]{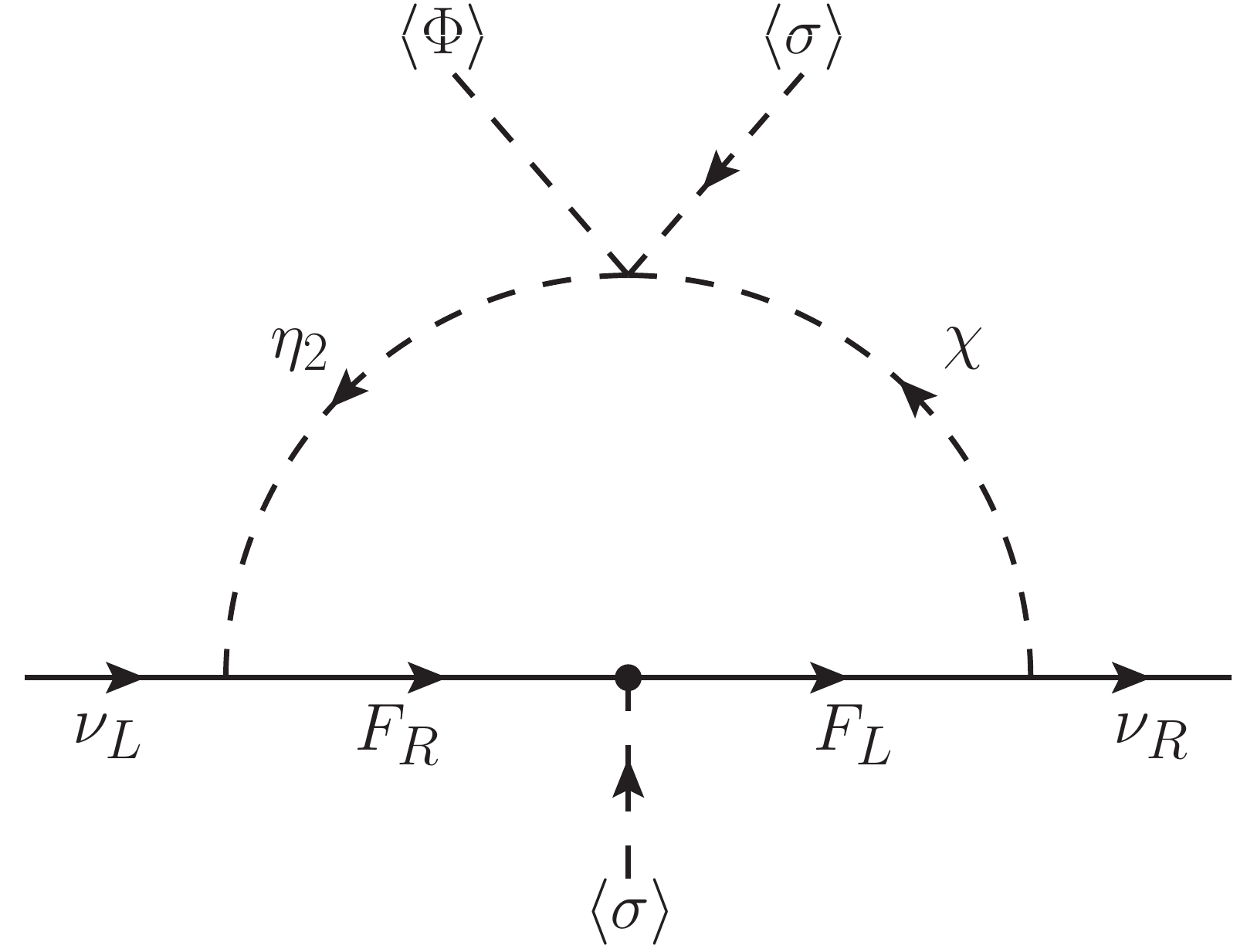}
\end{center}
\caption{ Dirac neutrino mass generation \label{nm}}
\end{figure}

The Yukawa sector for Dirac neutrino mass generation is given by
\begin{equation}
\mathcal{L}_{2} ~\supset~ h_{1}\bar{L}F_{R}\widetilde{\eta}_2+h_{2}\bar{\nu}_{R}F_{L}\chi^{\dagger}+f \bar{F}_{L}F_{R}\sigma+\hc
\label{y2}
\end{equation}
The effective mass matrix for active
neutrinos depicted in Fig. \ref{nm} is expressed as
\begin{equation}
(m_{\nu})_{\alpha\beta}=\frac{\sin2\theta}{8\pi^{2}}\sum_{k=1}^{6}
h_{1}^{\alpha k}m_{F_k}h_{2}^{k\beta}\Big[\frac{m_{S_{1}}^{2}}{m_{F_k}^{2}\!-m_{S_{1}}^{2}}
\log\big(\frac{m_{S_{1}}^{2}}{m_{k}^{2}}\big)-\frac{m_{S_{2}}^{2}}{m_{F_k}^{2}\!-m_{S_{2}}^{2}}
\log\big(\frac{m_{S_{2}}^{2}}{m_{F_k}^{2}}\big)\Big]\\
\end{equation}
where $m_{F_k}(k=1-4)$ denote the masses inert Dirac fermions. Typically, $m_\nu\sim0.1~\eV$ can be realised with $\theta\sim10^{-3}$, $h_1\simeq h_2\sim10^{-4}$, $m_F\sim1.5~\TeV$ and $m_{S_1,S_2}\sim10~\TeV$.
From Eq.\eqref{sc}, \eqref{y1} and \eqref{y2} , one can confirm that after the symmetry breaking with $v_{\phi}$ and $v_{\sigma}$ there exists a residual $Z_{2}$ symmetry for which the irrational $U(1)_X$ charged particles ($F_{Ri/Li}$, $\eta_{1}$ and $\chi$) are odd while other are even.  Therefore the lightest particles with irrational charges can not decay into SM particles and thus can be regarded as DM candidate.

\subsection{Lepton Flavor Violation}

The new Yukawa interactions of charged lepton will induce lepton flavor violation processes at one-loop level. Taking the radiative decay $\ell_\alpha\to \ell_\beta \gamma$ for an illustration, the corresponding branching ratio is calculated as \cite{Ding:2014nga}
\begin{eqnarray}
\text{BR}(\ell_\alpha\to\ell_\beta\gamma) &=&  \frac{3\alpha_\text{em}}{64\pi G_F^2}\left\{\left|
\sum_{i=1}^9\frac{y_1^{i\beta*}y_1^{i\alpha}}{m_{\eta_1^0}^2}
F_1\left(\frac{m_{\Psi_i}^2}{m_{\eta_1^0}^2}\right)+
\sum_{i=1}^4 \frac{h_1^{i\beta*}h_1^{i\alpha}}{m_{\eta_2^+}^2}
F_2\left(\frac{m_{F_i}^2}{m_{\eta_2^+}^2}\right)\right|^2\right. \\ \nonumber
&&\left.+\left|\sum_{i=1}^9 \sum_{\phi=h,H}
\frac{y_2^{i\beta*}y_2^{i\alpha}}{m_{\phi}^2}C^2_\phi
F_1\left(\frac{m_{\Psi_i}^2}{m_{\phi}^2}\right)
\right|^2\right\}\times\text{BR}(\ell_\alpha\to \ell_\beta \nu_\alpha \bar{\nu}_\beta),
\end{eqnarray}
where $C_h=\sin\alpha$ and $C_H=\cos\alpha$. And the loop functions $F_{1,2}(x)$ are given by
\begin{eqnarray}
F_1(x) &=& - \frac{2+3x-6x^2+x^3+6x\ln x}{6(1-x)^4}, \\
F_2(x) &=& \frac{1-6x+3x^2+2x^3-6x^2\ln x}{6(1-x)^4}.
\end{eqnarray}
According to previous discussion, $y_{1,2}^e=8.5\times10^{-4}$, $y_{1,2}^\mu=1.2\times10^{-2}$, $y_{1,2}^\tau=5.0\times10^{-2}$ with $m_{\Psi,\eta_1^0,H}\sim500~\GeV$ and $h_1\sim10^{-4}$ with $m_F\sim1.5~\TeV$, $m_{\eta_2^+}\sim10~\TeV$ are taken to reproduce lepton masses. Due to small Yukawa coupling $h$ and heavy mass of $\eta_2^\pm$, contribution of charged scalar $\eta_2^\pm$ is suppressed. The predicted branching ratios are $\text{BR}(\mu\to e\gamma)\simeq 1.6\times10^{-15}$, $\text{BR}(\tau\to e\gamma)\simeq 4.6\times10^{-15}$ and $\text{BR}(\tau\to \mu\gamma)\simeq 9.4\times10^{-13}$, which are clearly below current experimental limits \cite{TheMEG:2016wtm}.

\subsection{Mixing in the Gauge Sector}

Since $\eta_1$ is charged under both $U(1)_Y$ and $U(1)_X$, its vev $u$ will induce mixing between $Z_0$ and $Z_0'$ at tree level. The resulting mass matrix in the $(Z_0, Z_0')$ basis is given by\cite{Langacker:2008yv}
\begin{equation}
M^2=\left(\begin{array}{cc}
 \frac{1}{4} g_Z^2(v_\phi^2+u^2)&\frac{n}{2} g_Zg' u^2\\
  \frac{n}{2} g_Zg' u^2& g'^2 n^2 (v_\sigma^2 + u^2)
  \end{array}\right).
\end{equation}
The eigenvalues of $M^2$ are
\begin{equation}
m^2_{Z,Z'}=\frac{1}{2}\left[M^2_{11}+M^2_{22}\mp\sqrt{(M^2_{11}-M^2_{22})^2+4M^4_{12}}\right],
\end{equation}
with mixing angle given by
\begin{equation}
\tan2\theta_Z = \frac{2 M^2_{12}}{M^2_{22}-M^2_{11}}.
\end{equation}
As $u^2\ll v_{\phi}^2\ll v_{\sigma}^2$ in this model, we have $m_Z^2\simeq g_Z^2 v_\phi^2/4$, $m^2_{Z'}\simeq g'^2 n^2 v_\sigma^2$, and the mixing angle $\theta_Z\sim u^2/v_\sigma^2$ is naturally suppressed. Typically, for $u\sim10~\GeV$ and $v_\sigma\sim10~\TeV$, we have $\theta_Z\sim10^{-6}$.
Therefore, the dilepton signature $pp\to Z'\to \ell^+\ell^-$ at LHC is dramatically suppressed by the tiny mixing angle $\theta_Z$. For light $Z'$ around EW-scale, the four lepton signature $pp\to \ell^+\ell^- Z' \to \ell^+\ell^- \ell^+\ell^-$ is promising at LHC \cite{Bell:2014tta}. As shown in next section, the DAMPE excess favors heavy $Z'\lesssim3~\TeV$. In this case, the $Z'$ can hardly be detected at LHC, but are within the reach of the $3~\TeV$ CLIC in the $e^+e^-\to Z'\to \mu^+\mu^-$ channel \cite{Kara:2011xw}.

\subsection{LHC Signature}

In this subsection, we qualitatively discuss possible signatures of new particles at LHC. Since decays of $\eta_1$ scalars and $\Psi$ depend on their masses, the resulting signatures would be different. Considering the mass spectrum $m_\Psi<m_{\eta_1}$, the decay mode of $\eta_1$ scalars are $\eta_1^0\to \ell^-\Psi^+$, $\eta^\pm \to \nu\Psi^\pm$, and decay modes of $\Psi$ are
$\Psi^\pm\to \ell^\pm Z, \ell^\pm h, \nu W^\pm$. The promising signature would be $pp\to \Psi^+ \Psi^- \to \ell^+\ell^- ZZ$, leading to same signature as charged fermion in type-III seesaw \cite{Li:2009mw}. In the opposite case $m_\Psi>m_{\eta_1}$, the decay mode of $\eta_1$ scalars are $\eta_1^0\to \ell^+\ell^-$, $\eta_1^\pm\to \ell^\pm \nu$, and new decay modes of exotic charged fermion $\Psi^\pm \to \ell^\pm\eta_1^0, \nu \eta_1^\pm$ are also possible. Note that $\eta_1$ scalars are responsible for charged fermion mass, hence $\eta_1^0\to \tau^+\tau^-$ and $\eta_1^\pm\to \tau^\pm \nu$ are the dominant decay mode. The promising signature would be $pp\to \eta_1^0\eta_1^{0*}\to \tau^+\tau^-\tau^+\tau^-$, $pp\to \eta_1^\pm \eta_1^0\to \tau^\pm \nu \tau^+\tau^-$, similar as the lepton-specific 2HDM \cite{Abe:2015oca}.

For the mass of scalar singlet $m_H\sim500~\GeV$ with not too small mixing angle $\alpha\sim0.1$, the promising signature would be $gg\to H\to W^+W^-, ZZ, hh$ at LHC \cite{Robens:2016xkb}. Provided $m_\Psi<m_{H,\eta_1}$, the new decay channel $H\to \ell^\pm \Psi^\mp$ is also allowed. Then, the new signature $gg\to H\to \ell^\pm \Psi^\mp$ with $\Psi^\pm$ further decaying into $\ell^\pm Z,\nu W^\pm$ is a good way to probe the corresponding Yukawa coupling $y_2 \bar{E}_{R}\Psi_{L}\sigma$ introduced in this model.

As for the inert scalars, the most promising signature in principle would be $pp\to \eta_2^+\eta_2^-\to \ell^+ F_1 + \ell^- \bar{F}_1$, i.e., $\ell^+\ell^-+\cancel{E}_T$, for fermion DM at LHC \cite{Ding:2016wbd}. But actually, this dilepton signature is suppressed dramatically by heavy mass of the inert charge scalar $m_{\eta_2^\pm}\sim10~\TeV$ in our consideration \cite{Guella:2016dwo}, thus it is hard to probe at LHC. Similarly, the mono-$j$ signature $pp\to \eta_2^0 \eta_2^{0*} j \to \nu \bar{\nu} F_1 \bar{F}_1 j$, i.e., $j+\cancel{E}_T$, is also challenging at LHC.

\section{DAMPE Dark Matter}\label{Sec:DP}

Motivated by recent DAMPE excess around $1.5~\TeV$, we focus on DM phenomenon in this section. Here, we consider the lightest Dirac fermion $F_1$ as DM candidate. The relevant interactions mediated by the new gauge boson $Z'$ for DM and leptons are
\begin{equation}
\mathcal{L}_{Z'} \supset g'  Z^\prime_\mu \big( Q_{E_R} \bar{E}_R \gamma^\mu E_R +  Q_{\nu_R} \bar{\nu}_R \gamma^\mu \nu_R + Q_{F_L} \bar{F}_{L1} \gamma^\mu F_{L1} +  Q_{F_R} \bar{F}_{R1} \gamma^\mu F_{R1} \big),
\end{equation}
with mass of gauge boson $Z'$ given by $m_{Z'}\simeq g' n v_\sigma$.
In the following numerical calculation, we will take $n=1/3$ for illustration. Therefore, we have $Q_{E_R}=1$, $Q_{\nu_R}=2/3$, $Q_{F_L}=(\sqrt{11}+1)/6$, and $Q_{F_R}=(\sqrt{11}-1)/6$.

The dominant annihilation channels for DM $F_1$ are
\begin{equation}
\bar{F}_1 F_1 \to \bar{\ell}\ell, \bar{\nu}\nu, Z'Z'.
\end{equation}
Provided $m_{Z'}>m_{F_1}$, then the annihilation channel $\bar{F}_1 F_1\to Z'Z'$ is not allowed kinematically. Hence, $\bar{F}_1 F_1 \to \bar{\ell}\ell, \bar{\nu}\nu$ become dominant, which would be able to interpret the DAMPE $e^+ + e^-$ excess when $m_{F_1}\sim1.5~\TeV$.

\subsection{Constraints}

In this part, we summarize some relevant constraints for DAMPE DM. To research the DM phenomenon, we implement this model into {\tt FeynRules} \cite{Alloul:2013bka} package.  Then, for DM relic density, we require the results calculated by {\tt micrOMEGAs4.3.5} \cite{Belanger:2014vza} in $1\sigma$ range of Planck measurements: $\Omega h^2=0.1199\pm0.0027$ \cite{Ade:2015xua}.

As for direct detection, the leptophilic $Z'$ will mediate DM-electron scattering at tree level, with the corresponding cross section constrained by XENON100, i.e., $\sigma_e<10^{-34}\text{cm}^2$\cite{Aprile:2015ade}. Because of XENON100 sensitive to axial-vector couplings, the analytical expression for axial-vector DM-electron scattering is given by
\cite{Kopp:2009et}
\begin{equation}
\sigma_e = 3 (g_F^a g_\ell^a)^2 \frac{m_e^2}{\pi m_{Z'}^4}\approx 3 (g_F^a g_\ell^a)^2
\left(\frac{m_{Z'}}{10~\GeV}\right)^{-4}\times3.1\times10^{-39} \text{cm}^2,
\end{equation}
where $g_F^a=g'(Q_{F_R}-Q_{F_L})/2=-g'/6$ and $g_\ell^a=g' Q_{E_R}/2 = g'/2$. For $g'\sim0.1$, $m_{Z'}\sim3~\TeV$, the predicted value is far below current experimental bound. Instead, we consider the loop induced DM-nucleus scattering with the cross section calculated as \cite{Kopp:2009et}
\begin{equation}
\sigma_N = \frac{\alpha^2 Z^2 \mu_N^2}{9\pi^3 A^2 m_{Z'}^4} \sum_{\ell=e,\mu,\tau}
\left(g_F^v g_\ell^v \log \frac{m_\ell^2}{\mu^2} \right)^2,
\end{equation}
where $\mu_N= m_N m_{F_1}/(m_N+m_{F_1})$ is the reduced DM-nucleus mass, $g_F^v= g'(Q_{F_R}+Q_{F_L})/2 = g' \sqrt{11}/6$, $g_\ell^v=g' Q_{E_R}/2 = g'/2$ and $\mu=m_{Z'}/\sqrt{g_F^v g_\ell^v}$ is the cut-off scale. Since current most strict direct detection constraint is performed by PandaX~\cite{Cui:2017nnn}, we take $Z=54$, $A=131$ and $m_N=131~\GeV$ for the target nucleus¡¯ charge, mass number and mass respectively.

The leptophilic $Z'$ will contribute to anomalous magnetic moments of leptons \cite{Fayet:2007ua}
\begin{equation}
\Delta a_\ell \simeq \frac{g'^2}{12\pi^2} \frac{m_\ell^2}{m_{Z'}^2}
\end{equation}
For an universal gauge-lepton coupling, the precise measurement of $\Delta a_\mu = (27.8\pm8.8)\times10^{-10}$ \cite{Bennett:2006fi} set a stringent bound, i.e.,
$g'\lesssim 5\times10^{-3} m_{Z'}/ 1\GeV$. Meanwhile, searches for leptophilic $Z'$ at LEP in terms of four-fermion operators provide a much more stringent bound: $g'\lesssim 2\times10^{-4} m_{Z'}/1\GeV$ \cite{LEP:2003aa}.

\subsection{Fitting the DAMPE Excess}

\begin{figure}[!htbp]
\begin{center}
\includegraphics[width=0.45\linewidth]{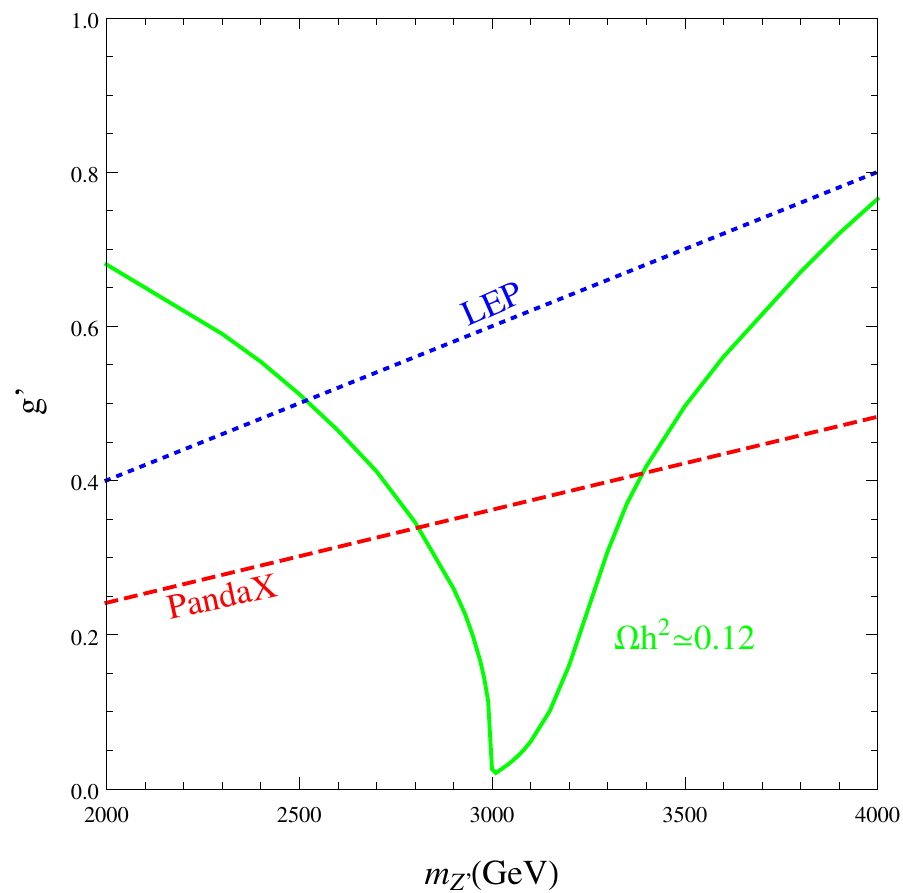}
\includegraphics[width=0.46\linewidth]{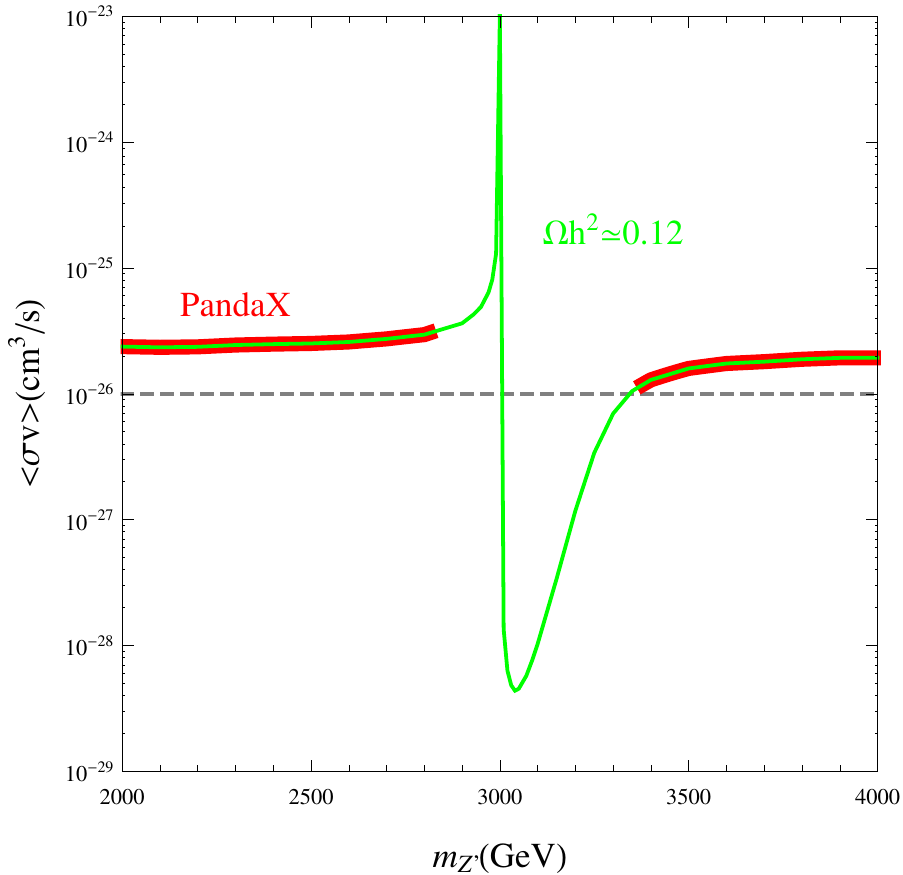}
\end{center}
\caption{ Left: Allowed region for DAMPE DM in the $g'\text{-}m_{Z'}$ plane. The green line delimit the relic density in the $1\sigma$ range: $\Omega h^2=0.1199\pm0.0027$, while blue and red line correspond to LEP and PandaX bound respectively. Right: Predicted value of current $\langle \sigma v \rangle$ in the halo as a function of $m_{Z'}$. The green line satisfy the observed relic density, while the red curves are excluded by PandaX.}
\label{FIG:gp}
\end{figure}

To determine the allowed parameter space under above constraints from relic density, direct detection and collider searches, we scan over the $g'\text{-}m_{Z'}$ plane while fix $m_{F_1}=1500~\GeV$. The results are depicted in Fig.~\ref{FIG:gp}. Since the dominant annihilation channels into leptons are via $s$-channel, the resonance production of $Z'$ will diminish the required $g'$ coupling for correct relic density. And currently, the most stringent bound is from direct detection, which constrains $Z'$ around the resonance region. In Fig.~\ref{FIG:gp}, the predicted value of current $\langle \sigma v \rangle$ in the halo is also shown. Slightly below the resonance, the Breit-Wigner mechanism \cite{Ibe:2008ye} greatly enhances the annihilation cross section. In contrast, we see a strong dip just above the resonance. Considering the fact that DAMPE excess favor $\langle \sigma v \rangle>10^{-26}\text{cm}^3/\text{s}$ as well as PandaX has excluded the region $m_{Z'}<2810~\GeV\cup m_{Z'}>3380~\GeV$, the possible region to interpret DAMPE excess falls in the range $m_{Z'}\in[2810,3000]~\GeV$.

\begin{table*}[hbtp]
\begin{tabular}{|c|c|c||c|c|c||c|c|}
\hline
$M_{\rm DM}$ (GeV) & $M_{Z'}$ (GeV) & $g^\prime$ & $\Omega_{\rm DM} h^2$ & $\langle\sigma v\rangle$ (${\rm cm}^3/{\rm s}$) & $\sigma_{\rm SI}$ (${\rm cm}^2$) & $\ell\bar{\ell}$ & $\nu_{\ell}\bar{\nu}_{\ell}$  \\\hline
$1500$ & $2950$ & $0.2$ & $0.1192$  & $4.9\times10^{-26}$ & $1.8\times10^{-46}$ & $70\%$ & $30\%$ \\\hline
\end{tabular}
\caption{The DM information for benchmark point to fit DAMPE excess. Here $\langle\sigma v\rangle$ is thermal averaged cross section at present. The last two columns present relative contributions for various annihilation channels.}
\label{tab:bench}
\end{table*}

\begin{figure}[!htbp]
\begin{center}
\includegraphics[width=0.8\linewidth]{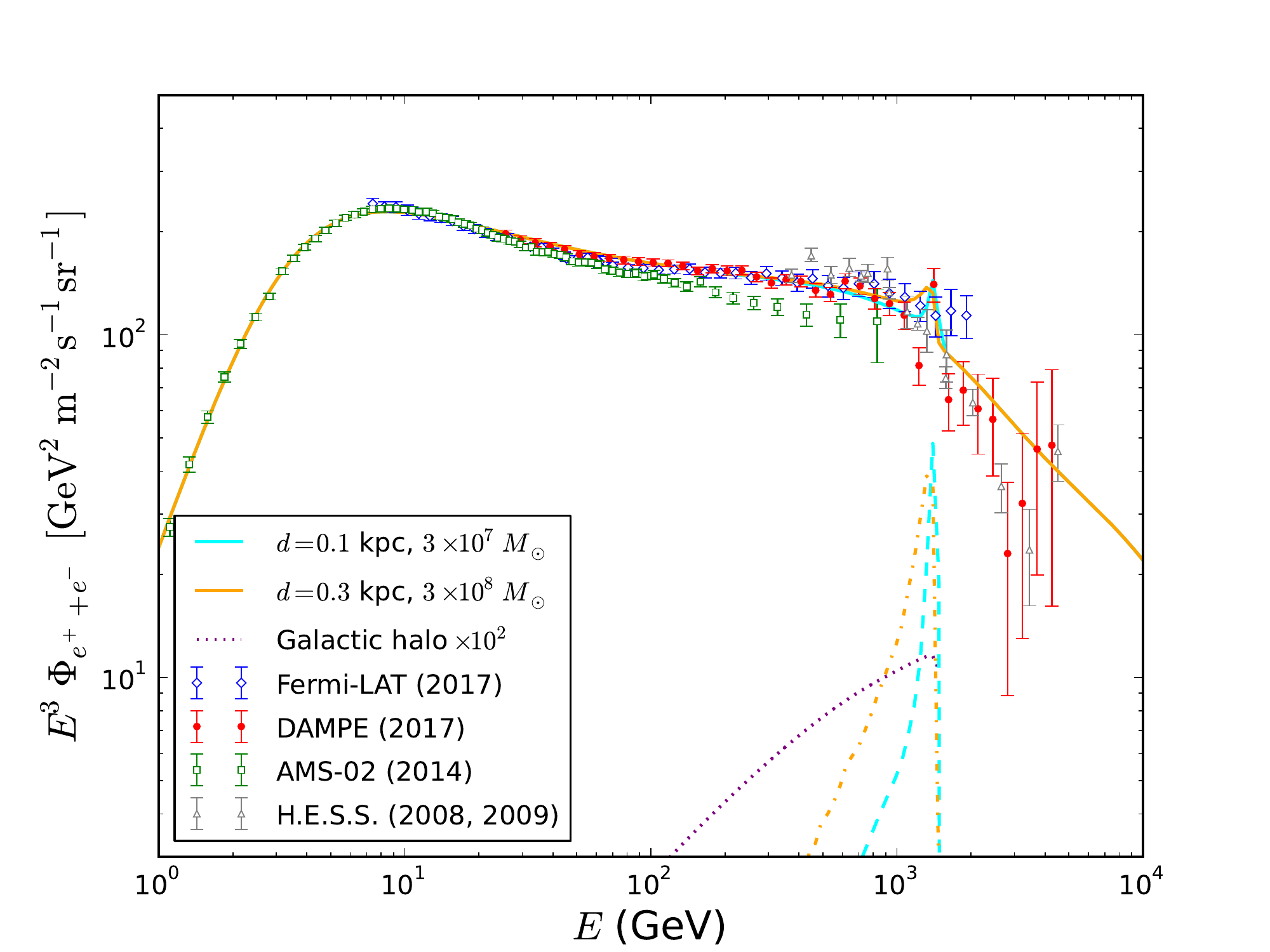}
\end{center}
\caption{The $e^+ + e^-$ flux for our benchmark point. The DAMPE data is shown in red points \cite{Ambrosi:2017wek}. The blue dotted line corresponds to Galactic halo contribution (multiplied by $10^2$), and cyan dashed (orange dot-dashed) line corresponds to contribution of nearby subhalo with a distance of $0.1$ ($0.3$) kpc and the viral mass $3\times 10^7$ $M_{\odot}$ ($3\times 10^8$ $M_{\odot}$). Corresponding total fluxes (background + Galactic halo+ subhalo) are also shown by solid lines with the same colors. Here we take solar modulation potential as $\Phi_{\odot}=700$ MV for illustration. For comparison, the direct measurements from AMS-02~\cite{Aguilar:2014mma} and Fermi-LAT~\cite{Abdollahi:2017nat} experiments, as well as the indirect measurement by H.E.S.S.~\cite{Aharonian:2008aa,Aharonian:2009ah} are also shown. The error bars of DAMPE, AMS-02 and Fermi-LAT include both systematic and statistical uncertainties.}
\label{fig:bench}
\end{figure}

Based on the above analysis, we select a benchmark point (see Table~\ref{tab:bench}) to fit the sharp DAMPE excess by taking into account contributions from both nearby subhalo and Galactic halo. In our numerical calculation, we respectively use {\tt GALPROP}~\cite{Moskalenko:1997gh,Strong:1998pw} and {\tt micrOMEGAs} packages \cite{Belanger:2014vza} to evaluate the background flux coming from various astrophysical sources and the flux due to DM annihilation in Galactic halo. While for subhalo contribution, we numerically solve following steady-state diffusion equation~\cite{Cirelli:2010xx}
\begin{align}
-\vec{\nabla}\cdot\left[K(E)\vec{\nabla}f(\vec{x},E)\right]-\frac{\partial}{\partial E}\left[ b(E)f(\vec{x},E)\right]=Q(\vec{x},E)\,,
\label{eq:diffusion}
\end{align}
with the source term
\begin{align}
Q(\vec{x},E) = \frac{\langle \sigma v \rangle}{2m_{\rm DM}^2}\int \rho^2(r)dV \delta^3(\vec{x} -\vec{x}_{\rm sub})\;,
\label{eq:source}
\end{align}
by using Green function method. In Eqs.~(\ref{eq:diffusion}) and~(\ref{eq:source}),
$K(E)=K_0(E/E_0)^\delta$ is the diffusion coefficient, $b(E)=E^2/(E_0\tau_E)$ is the positron loss rate due to the synchrotron radiation and inverse Compton scattering, $\langle \sigma v \rangle$ the thermal averaged cross section at present, $\rho(r)$ and $x_{\rm sub}$ the density profile and location of nearby subhalo, respectively. Here we adopt propagation parameters as~\cite{Yuan:2017ysv}: $K_0=0.1093$ ${\rm kpc}^2$ ${\rm Myr}^{-1}$, $\delta=1/3$, $L=4$ kpc (the half height of the Galactic diffusion cylinder), $\tau_E=10^{16}$ $s$ (the typical loss time) and $E_0=1$ GeV. In addition, we assume both Galactic halo and subhalo are follow NEW density profile~\cite{Navarro:1995iw,Navarro:1996gj}:
\begin{align}
\rho(r)=\frac{\rho_s}{(r/r_s)(1+r/r_s)^2}\,.
\end{align}
The Galactic halo is normalized by the local density $\rho_{\odot}$ at Sun orbit $R_{\odot}$, which are respectively fixed as $\rho_{\odot}=0.4$ ${\rm GeV}{\rm cm}^{-3}$ and $R_{\odot}=8.5$ kpc. While for nearby subhalo, the parameters $\rho_s$ and $r_s$ can be determined by its viral mass $M_{\rm vir}$. The fitting result for our benchmark point is presented in Fig.~\ref{fig:bench} together with DAMPE data points. From which, we find that a nearby subhalo with a distance of $0.1$ ($0.3$) kpc and the viral mass $3\times 10^7$ $M_{\odot}$ ($3\times 10^8$ $M_{\odot}$) can account for the DAMPE excess for our model.

\section{Conclusion}\label{Sec:CL}

In this paper, we propose an anomaly-free radiative seesaw model with an alternative leptophilic $U(1)_X$ gauge symmetry. Under the $U(1)_X$ symmetry, only right-handed leptons are charged. Charged leptons acquire mass via the type-I seesaw-like mechanism with heavy intermediate fermions added also for anomaly-free cancellation. Meanwhile, tiny neutrino masses are generated at one-loop level with DM candidate in the loop.

Provided all other particles are heavy enough, the dominant annihilation channel for DM $F_1$ is $\bar{F}_1F_1\to \bar{\ell}\ell,\bar{\nu}\nu$ mediated by the new leptophilic gauge boson $Z'$.
Motivated by the observed DAMPE $e^++e^-$ excess around $1.5~\TeV$, we fix $m_{F_1}=1.5~\TeV$ while consider possible constraints from relic density, direct detection and collider searches. Under all these constraints, a benchmark points, i.e., $m_{Z'}=2950~\GeV$, is chosen from the viable region $m_{Z'}\in [2810,3000]~\GeV$. After fitting to the observed spectrum,  we find that the DAMPE excess can be explained by a nearby subhalo with a distance of $0.1$ ($0.3$) kpc and the viral mass $3\times 10^7$ $M_{\odot}$ ($3\times 10^8$ $M_{\odot}$).

\section*{Acknowledgements}
The work of Weijian Wang is supported by National Natural Science Foundation of China under Grant
Numbers 11505062, Special Fund of Theoretical Physics under Grant Numbers 11447117 and Fundamental
Research Funds for the Central Universities under Grant Numbers 2014ZD42.
We thank Qiang Yuan for help on DAMPE spectrum fitting.

\end{document}